\documentclass[useAMS,usegraphicx,usenatbib]{mn2e}
\usepackage{amssymb}
\usepackage{times}

\title[The surface mass densities of CDM haloes]{An analytical model of surface mass densities of cold dark matter haloes -- with an application to MACHO microlensing optical depths}

\author[J. Holopainen et al.]{Janne Holopainen$^{1}$, Erik Zackrisson$^{1,2,3}$, Alexander Knebe$^{4}$, Pasi Nurmi$^{1}$, \newauthor Pekka Hein\"am\"aki$^{1}$, Chris Flynn$^{1}$, Stuart Gill$^{5}$ and Teresa Riehm$^{2}$\\
$^1$Tuorla Observatory, University of Turku, V\"ais\"al\"antie 20, Piikki\"o, FI-21500, Finland\\
$^2$Stockholm Observatory, AlbaNova University Center, 106 91 Stockholm, Sweden\\
$^3$Department of Astronomy and Space Physics, Box 515, SE-75120 Uppsala, Sweden\\
$^4$Astrophysikalisches Institut Potsdam, An der Sternwarte 16, 14482 Potsdam, Germany \\
$^5$Department of Astronomy, Columbia University, 550 West 120th Street, New York, NY 10027, USA
}
\begin{document}
\date{Accepted ... Received ...; in original form ...}
\pagerange{\pageref{firstpage}--\pageref{lastpage}} \pubyear{2007}
\maketitle
\label{firstpage}

\begin{abstract}
The cold dark matter (CDM) scenario generically predicts the existence of triaxial dark matter haloes which contain notable amounts of substructure. However, analytical halo models with smooth, spherically symmetric density profiles are routinely adopted in the modelling of light propagation effects through such objects. In this paper, we address the biases introduced by this procedure by comparing the surface mass densities of actual N-body haloes against the widely used analytical model suggested by Navarro, Frenk and White (1996) (NFW). We conduct our analysis in the redshift range of 0.0 - 1.5.

In cluster sized haloes, we find that triaxiality can cause scatter in the surface mass density of the haloes up to $\sigma_+ = +60 \%$ and $\sigma_- = -70 \%$, where the 1-$\sigma$ limits are relative to the analytical NFW model given value. Subhaloes can increase this scatter to $\sigma_+ = +70 \%$ and $\sigma_- = -80 \%$. In galaxy sized haloes, the triaxial scatter can be as high as $\sigma_+ = +80 \%$ and $\sigma_- = -70 \%$, and with subhaloes the values can change to $\sigma_+ = +40 \%$ and $\sigma_- = -80 \%$.

We present an analytical model for the surface mass density scatter as a function of distance to the halo centre, halo redshift and halo mass. The analytical description enables one to investigate the reliability of results obtained with simplified halo models. Additionally, it provides the means to add simulated surface density scatter to analytical density profiles. As an example, we discuss the impact of our results on the calculation of microlensing optical depths for MACHOs in CDM haloes.

\end{abstract}
\maketitle

\section{Introduction}

The cold dark matter model, in which the non-baryonic part of the dark matter is assumed to consist of particles that were non-relativistic already at the time of decoupling, and that interact predominantly through gravity, has been very successful in explaining the formation of large-scale structures in the Universe \citep[see e.g.][ for a review]{Primack}. In this scenario, both galaxies and galaxy clusters are hosted by CDM haloes, which formed hierarchically through mergers of smaller subunits. 

Even though N-body simulations generically predict CDM haloes to be triaxial \citep[e.g.][]{Jing & Suto} with substantial amounts of substructures left over from the merging process \citep[e.g.][]{Moore et al.}, simplified halo models are often adopted in the modelling of light propagation through such objects. The most common approach is to treat dark matter haloes as spherical objects with smooth density profiles, usually either of the NFW \citep{NFW} form, some generalization thereof \citep{Zhao}, or that of a cored or singular isothermal sphere.

The light emitted from high-redshift objects such as quasars, supernovae, gamma-ray bursts, galaxies and galaxy clusters will typically have to pass through many dark matter haloes before reaching an observer on Earth. Several investigations have already indicated that smooth and/or spherical halo models may lead to incorrect results when treating the gravitational lensing effects associated with such foreground mass condensations \citep[e.g][]{Bartelmann & Weiss,Dalal et al.,Oguri & Keeton,Hennawi et al.}

More realistic features like triaxiality and substructures can be included in gravitational lens calculations either by employing N-body simulations directly \citep[e.g.][]{Bartelmann & Weiss,Seljak & Holz,Holopainen et al.} or by using analytical expressions for the halo shapes \citep[e.g.][]{Kochanek,Golse & Kneib,Evans & Hunter,Chae} and subhalo properties \citep[e.g.][]{Oguri b,Zackrisson & Riehm}. While N-body simulations often represent the safest choice, the approach is computationally demanding and does not always allow one to identify the features of the mass distribution responsible for a specific lensing effect. Methods which bring simple, analytical halo models into contact with the full phenomenology of the N-body simulations are therefore highly desirable. 

In this paper, we focus on the projected mass density of CDM haloes as a function of distance from the halo centre. There are several situations in gravitational lensing when realistic estimates of the surface mass density (i.e. convergence) along a given line of sight through a dark halo may be important. Examples include the calculation of image separations in strong lensing by subhaloes located in the external potential of its host halo \citep{Oguri b}, attempts to correct the luminosities of supernovae type Ia for the magnification by foreground haloes \citep[e.g.][]{Gunnarsson} and estimates of the distribution of microlensing optical depths for high-redshift MACHOs \citep[e.g.][]{Wyithe & Turner,Zackrisson & Riehm}. Other applications include the assessments of light propagation effects in models with non-zero coupling betweeen dark matter particles and photons \citep[e.g.][]{Profumo & Sigurdson}.

Here, we use high-resolution, dissipationless N-body simulations of CDM haloes to investigate the errors in surface mass density introduced by treating these objects as spherical with smooth density profiles of the NFW type. Simple relations for the surface mass density error as a function of halo redshift and distance to the halo centre are presented, making it easy to investigate the reliablitiy of results obtained with simplified halo models.  

On a related note, \citet{KnebeWiessner} recently investigated the error introduced by spherically averaging an elliptical mass distribution. They found that for axis ratios typical for cosmological dark matter haloes, the variance in the local density can be as large as 50\% in the outer parts. The current paper examines the problem of halo triaxiality from a slightly different point of view.

The N-body simulations used are described in Section 2. In Section 3, we describe the methods for extracting the halo sample. In Section 4, we compare the CDM surface mass densities obtained along random sightlines through the N-body haloes to the corresponding results obtained from smooth and spherical NFW models fitted to the same haloes. Section 5 presents a set of simple relations for the surface mass density errors introduced by this procedure as a function of distance to the halo centre and halo redshift. Section 6 discusses how these relations may be used in the context of optical depth estimates for MACHO microlensing. A number of caveats are discussed in Section 7. Section 8 summarizes our findings.

\section{N-body simulations}

\begin{flushleft}
\begin{figure}
\includegraphics[height=84mm, angle=-90]{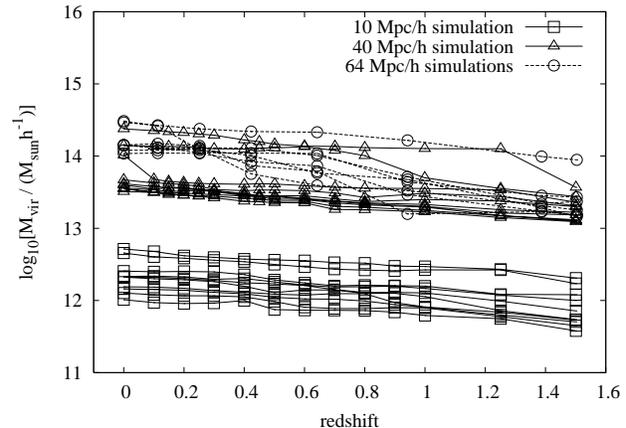}
\caption{
The masses and redshifts of the haloes used in our analysis. The total number of haloes is 336. See text for details.
}
\label{halo_masses}
\end{figure}
\end{flushleft}

\subsection{The 64 $h^{-1}$Mpc simulations}

For studying cluster sized haloes, we utilize a suite of {\it four} high-resolution N-body simulations. The simulations were carried out using the publicly available adaptive mesh refinement code \texttt{MLAPM} \citep{MLAPM}, focusing on the formation and evolution of dark matter galaxy clusters containing of order one million particles, with mass resolution $1.6 \times 10^8$ $h^{-1}{\ }{\rm M_{\odot}}$ and spatial force resolution $\sim$2$h^{-1}{\ }{\rm kpc}$. They are so-called ``zoom'' or multimass simulations in which we first created a set of four independent initial conditions at redshift $z=45$ in a standard $\Lambda$CDM cosmology ($\Omega_0 = 0.3,\Omega_\lambda = 0.7, \Omega_b = 0.04, h = 0.7, \sigma_8 = 0.9$). $512^{3}$ particles were placed in a box of side length 64$h^{-1}{\ }{\rm Mpc}$ giving a mass resolution of $m_p = 1.6 \times 10^{8}$$h^{-1}{\ }{\rm M_{\odot}}$.  For each of these initial conditions we iteratively collapsed eight adjacent particles to a single particle, reducing our simulation to 128$^3$ particles. These lower mass resolution initial conditions were then evolved until $z=0$. At $z=0$, eight clusters from different regions of our simulations were selected: 4 halos from box \#1, one from box \#2, one from box \#3 and 2 from box \#4. The masses of these haloes are in the range  1--3$\times 10^{14}$$h^{-1}{\ }{\rm M_{\odot}}$ and triaxiality parameters vary from 0.1 to 0.9. Then, as described by \citet{Tormen97}, for each cluster the particles within five times the virial radius were tracked back to their Lagrangian positions at the initial redshift ($z=45$). Those particles were then regenerated to their original mass resolution and positions, with the next layer of surrounding large particles regenerated only to one level (i.e. 8 times the original mass resolution), and the remaining particles were left 64 times more massive than the particles resident with the host cluster. This conservative criterion was selected in order to minimise contamination of the final high-resolution haloes with massive particles.

A more elaborate description of this data set and a detailed investigation of the sense of rotation of the satellites and the properties of the tidally induced debris field of disrupting satellites can be found elsewhere \citep{WK, WKPI, WKPII}.

\subsection{The 10 $h^{-1}$Mpc and 40 $h^{-1}$Mpc simulations}

For studying galaxy sized halos ($M \sim 10^{12} \rm M_{\odot}$) and for acquiring better statistics on the larger haloes, we ran two additional simulations with smaller box sizes. The same simulation code \texttt{MLAPM} as in the 64 $h^{-1}$Mpc simulations was used, but the cosmological constants were slightly different: $\Omega_0=0.27$, $\Omega_b=0.044$ and $\Omega_{\lambda}=0.73$. These two simulations are standard cosmological simulations where all the particles have the same mass: $4.47\times10^6$$h^{-1}{\ }{\rm M_{\odot}}$ and $2.86\times10^8$$h^{-1}{\ }{\rm M_{\odot}}$ for 10 $h^{-1}$Mpc and 40 $h^{-1}$Mpc simulations, respectively. Spatial force resolutions were 0.46 $h^{-1}$kpc and 1.8 $h^{-1}$kpc. Both simulations were followed until $z=0$ and halos were identified at different redshifts. The 10 $h^{-1}$Mpc simulation was followed from $z=71.52$ and the 40 $h^{-1}$Mpc simulation from $z=47.96$.

One may argue that the large scale modes, ignored in these relatively small volume simulations, can cause spurious errors in the results. Especially in the smallest simulation volume, the long wavelength perturbations will not be present. However, the distributions of mass concentrations remain the same when compared to larger volume simulations. Also, in the 10 $h^{-1}$Mpc box, we restrict our analysis to intermediate mass halos and their subhalos. We believe that the simulations are reliable for our purposes and in agreement with simulations covering larger volumes.

\section{CDM Haloes}

\subsection{Finding and truncating haloes}

Finding and truncating dark matter haloes within cosmological simulations is an interesting and challenging task. Many authors have addressed this problem successfully by developing sophisticated algorithms which can locate haloes by a variety of techniques \citep[e.g.][]{Davis et al., Frenk et al., Bertschinger & Gelb, SCO, WHK, Klypin & Holtzman}.

Our analysis examines the cores as well as the outskirts of the haloes without the luxury of being allowed to overlook the exact properties of the low density regions. We need to study all parts or the halo density profile in three dimensions, and we need to get rid of the background particles at all radii as well as possible, especially near the virial radius. To achieve this, we use a highly capable halo finder which can determine the parent potentials of individual particles and ``clean'' host haloes from their subhaloes.

We find and truncate our haloes and subhaloes using the \texttt{MLAPM} Halo Finder (MHF) \citep{Gill et al.}. MHF uses the adaptive grids of \texttt{MLAPM} to locate haloes within the simulation. \texttt{MLAPM}'s adaptive refinement meshes follow the density distribution {\it by construction}. The grid structure naturally ``surrounds'' the haloes, as the haloes are simply manifestations of over-densities in the mass distribution of the simulation box. The grids of \texttt{MLAPM} are adaptive, and it constructs a series of embedded grids, the higher refinement grids being subsets of grids on lower refinement levels. MHF takes this hierarchy of nested isolated grids and constructs a ``grid tree''. Within that tree, each branch represents a halo, thus identifying haloes, subhaloes, subsubhaloes and so on. 

While a branch of the tree identifies the majority of particles associated with a halo, the surrounding region is checked for additional particles if the halo is embedded within another halo (i.e., it is a subhalo). To gather additional particles, a larger collection radius is defined, and all the particles within this radius are assigned to the halo. In this paper, the collection radius has been defined as half the distance between the current halo and the next most massive halo.

The gravitationally unbound particles are then removed from the haloes in an iterative fashion. If a particle is not bound, it is assigned to the subhalo's host or the background as appropriate. This, however, does not guarantee that each particle is uniquely assigned to a halo. It is possible for a particle to be shared by two or more halo potentials by these criteria.

Using this set of particles, the canonical properties of the haloes are calculated. For example the virial radius is found by stepping out in (logarithmically spaced) radial bins until the density reaches $\rho_{\rm halo}(r_{\rm vir}) = \Delta_{\rm vir}(z) \rho_b$(z), where $\rho_b$ is the cosmic matter density. Particles outside this radius are removed. If this density is not reached, then we consider the furthest bound particle from the centre of the halo as the radius.

\subsection{The halo sample}

Typically, MHF finds thousands of haloes within a simulation box, but the number of haloes which have sufficient mass resolution for our purposes is unfortunately quite low. When choosing the halo sample, we have to balance between particle resolution and the number of haloes. After looking at the consistency of the NFW fit and the halo mass profiles, we decided to include the ten most massive haloes from the 10 $h^{-1}$Mpc and 40 $h^{-1}$Mpc simulations. This choice translates to a minimum resolution of 36,000 particles in the smallest halo in our sample. Note that our particle counts are given without subhaloes, which normally contribute about $10 - 20$ \% of the host's mass.

Nearly all haloes in the sample contain more than 100,000 particles at the maximum redshift, $z_{\rm max} = 1.5$. The only exception is the 40 $h^{-1}$Mpc box, in which the number of particles contained by the tenth most massive halo is only 36,000. However, this is compensated by the large particle counts in the multimass simulations, which contain haloes in the same mass range. The particle counts of the halo sample is shown in Table \ref{npart_table}. Note that the largest number of particles within a halo is always found at $z = 0$ and the smallest number of particles at $z = 1.5$.

Each of the eight 64 $h^{-1}$Mpc multimass simulations contains only one high resolution cluster halo region. When the virial mass accretion histories were examined, we found that the youngest of the eight clusters is composed of three smaller haloes with comparable masses. In this case, it was impossible for MHF to separate the three interacting haloes from each other, and therefore, we were left with seven well defined haloes per redshift from the multimass simulations. Excluding the ``triple-cluster'' does not introduce any selection bias since this type of specific merger occurs rarely.

The number of redshifts used per simulation box can be seen in Fig. \ref{halo_masses}. The redshift counts are 13, 15 and 8 for the 10 $h^{-1}$Mpc, 40 $h^{-1}$Mpc and 64 $h^{-1}$Mpc boxes, respectively. In total, the number of haloes in the sample is 130 + 150 + 56 = 336. 

Our halo sample is divided into two mass classes with a mass gap at $M \sim 10^{13}$ M$_\odot h^{-1}$ at $z = 0$ (see Fig. \ref{halo_masses}). We refer to the more massive class as the {\it cluster sized} haloes and to the lower mass class as the {\it galaxy sized} haloes. We would need another cosmological simulation with a comoving box size of $\sim$ 25 $h^{-1}$Mpc to fill the mass gap. The 40 $h^{-1}$Mpc simulation does contain haloes with masses at the range of the gap, but unfortunately the resolutions of these haloes are insufficient for our purposes.

\begin{table}
\begin{center}
\caption{
Summary of the number of haloes and the particle counts in our sample. First column gives the simulation box size and $N_{\rm haloes}$ refers to the number of haloes the sample contains per redshift. $N_{\rm part}^{\rm min}$ refers to the number of particles within $r_{\rm vir}$ of the least massive halo and $N_{\rm part}^{\rm max}$ to the number of particles within $r_{\rm vir}$ of the most massive halo, in each box.
}
\begin{tabular}{cccccc}
\hline
box   & $N_{\rm haloes}$ & $N_{\rm part}^{\rm min}$ & $N_{\rm part}^{\rm max}$ & $N_{\rm part}^{\rm min}$ & $N_{\rm part}^{\rm max}$ \\
$h^{-1}$Mpc &                  & $z=1.5$                  & $z=1.5$                  & $z=0$                    & $z=0$ \\
\hline \hline
10 & 10 & 75,000 & 320,000 & 220,000 & 1,000,000 \\
40 & 10 & 36,000 &  93,000 & 100,000 &   760,000 \\
64 & 7  & 85,000 & 380,000 & 600,000 & 1,600,000 \\
\hline
\label{npart_table}
\end{tabular}
\end{center}
\end{table}

\subsection{Analytical haloes} \label{analytical_haloes}
Once we have found, extracted and chosen our N-body haloes, we want to construct their analytical counterparts. We have chosen to use the NFW density profile so that our fit procedure can be compared to earlier work and reproduced easily. The fit equation is
\begin{equation} \label{nfw}
\rho(r) = \frac{a_0}{\frac{r}{a_1}(1 + \frac{r}{a_1})^2},
\end{equation}
where $a_0$ and $a_1$ are our fit parameters. We measure the number density of the particles in a halo at a certain radius by dividing the particles in logarithmically spaced radial bins (shells) and calculating the particles within these bins. Then, by the fit procedure (Levenberg-Marquardt algorithm), we find the $a_0$ and $a_1$ values which minimize
\begin{equation} \label{nfw_chi}
\chi^2 = \sum_i \left(\frac{\rho(r_i) - \rho_{\rm s}(r_i)}{\sigma_{\rho}(r_i)}\right)^2,
\end{equation}
where $\rho_{\rm s}(r_i)$ is the measured number density of the $i$th shell and $r_i$ is the midpoint radius of that shell. The estimated ``measurement error'' in the number of particles within a shell is assumed to be Poissonian: $\sigma_{N_i} = \sqrt{N_i}$. This makes the estimated ``measurement error'' in the number density to be $\sigma_{\rho}(r_i) = N_i/\rho_{\rm s}^2(r_i)$. The choice of weighting the fit with $\sigma_{\rho}(r_i)^2$ tends to provide the inner core less weight than to the outer regions. This is a deliberate choice because our analysis is affected by the full structure of the halo and only in few rare cases dominated by the center.

\section{Measuring the surface mass density} \label{measuring_sigma}

\subsection{Sightlines}
After we have assigned the particles to a certain halo, we smooth the mass distribution by treating the particles as triangular-shaped clouds \citep{hockney}. This approach allows us to measure the surface mass densities of the haloes using one-dimensional sightlines. If the particles were treated as point masses, the surface mass density would have to be measured using tubes of finite radius, which would lead to an average measurement within a tube. To avoid the averaging process, each particle is assigned a surface mass density
\begin{equation} \label{surfacemass}
\Sigma_{\rm p}(r) =
\left\{
\begin{array}{ll}
\displaystyle \Sigma_0 (1 - \frac{r}{a})
= \frac{3 M_{\rm p}}{\pi a^3} (1 - \frac{r}{a}) & \mbox{$r < a$} \\
0  & \mbox{$r \geq a$}
\end{array}
\right.
\end{equation}
where $r$ is the impact parameter between a sightline and a particle and $M_{\rm p}$ is the mass of the particle. $a$ is the size of the adaptive grid cell where the particle is found, and $a$ is used in our analysis as the effective radius of the particle. The cell size depends from the local particle density and is determined by the cosmological simulation grid constructing process (MHF and \texttt{MLAPM} can both construct the same adaptive grid tree). The cell sizes in the simulations can range from $250$ $h^{-1}$kpc to $0.150$ $h^{-1}$kpc, and the smallest cells are automatically located in the dense cores of haloes and subhaloes.

To sample the surface mass density of an MHF extracted halo, we pierce it with 10,000 randomly oriented sightlines with impact parameters chosen from a uniform random distribution in the range $r_{\rm imp} \in (0, r_{\rm vir})$. We then calculate the surface mass density for each sightline $\Sigma_{\rm SL}(r_{\rm imp})$ by summing the surface mass densities of the particles for which $r < a$.

The proper truncation of the smooth mass distribution at the virial radius is slightly complicated. Some of the particles can extend a portion of their mass beyond the virial radius. For these particles, only the mass inside $r_{\rm vir}$ is taken into account. Furthermore, the extended particles introduce an additional requirement for our halo finder -- it needs to be able to find particles which belong to a given halo potential (based on the particle velocity) out to $\sim 1.5 r_{\rm vir}$. This is because most of the particles between $1.0 < r_{\rm vir} < 1.5$ extend their effective radii (defined in Eq. \ref{surfacemass}) into the region inside the virial radius and contribute mass to our sightlines. 

\subsection{Truncated density profile}
We want to analyse how the surface mass density of an N-body halo behaves compared to an analytical model. To calculate the analytical value, we use a truncated version of the projected NFW density profile, which gives us the predicted surface mass density as a function of the impact parameter ($r_{\rm imp}$) and the virial radius of a halo ($r_{\rm vir}$). Truncation is needed because our haloes have a limited size, unlike the NFW haloes, which extend to infinity. The fit parameters $a_0$ and $a_1$ are also used in the truncated model -- after they have been determined by fitting Eq. \ref{nfw} to the respective halo.

We derive the following form for the truncated NFW surface mass density
\begin{equation} \label{tnfw}
\Sigma_{\rm NFW}(x, c) = \int_{z_0}^{z_1} \rho(z) dz = 2 a_0 a_1 F(x, c),
\end{equation}
where
\begin{equation} \label{sigma_nfw}
F(x, c) =
\left\{
\begin{array}{ll} 
\frac{1}{x^2-1}  \left( \frac{\sqrt{c^2-x^2}}{1 + c} - \frac{{\rm cosh}^{-1}( \frac{c+x^2}{(1 + c)x}) }{\sqrt{1-x^2}} \right)
& \mbox{$x < 1$} \\
\frac{1}{x^2-1} \left( \frac{\sqrt{c^2-x^2}}{1 + c} - \frac{{\rm cos}^{-1}( \frac{c+x^2}{(1 + c)x} )}{\sqrt{x^2-1}} \right)
& \mbox{$x > 1.$}
\end{array}
\right.
\end{equation}
Here $x = r_{\rm imp}/a_1$ and $c = r_{\rm trunc}/a_1$. Note that $c$ is analogous to the NFW concentration parameter $c_{\rm vir}$ when the truncation radius is equal to the virial radius $r_{\rm vir}$. When $|x - 1| < 0.1$, we have to interpolate (linearly) between both forms of the function, $F_{x>1}(1.1x, c)$ and $F_{x<1}(0.9x, c)$ for numerical reasons.

\begin{figure}
\includegraphics[width=84mm, angle=0]{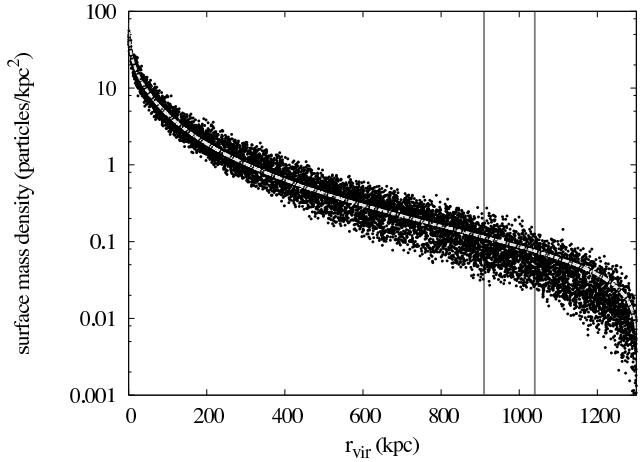}
\caption{
This figure shows how the surface mass density scatters due to triaxiality in most of the haloes in our halo sample. Here the values are measured from a cluster sized halo with 10,000 randomly oriented sightlines, {\it without} subhaloes. The dashed line shows the truncated NFW fit profile, and the two vertical lines mark the borders of a radial bin. The $\Sigma_{\rm SL}$-distribution of this bin is fitted with a log-normal function in Fig. \ref{lnfit}. 
}
\label{sigma_scatter}
\end{figure}

For the untruncated version of Eq. \ref{tnfw}, in which $c\rightarrow\infty$, see e.g. \cite{Golse & Kneib}. The untruncated model extends asymptotically to infinity and does not have the turnover seen for example in Figure \ref{sigma_scatter} between $800 {\rm kpc} < r < 1300 {\rm kpc}$. The surface mass density has also been derived by \citet{Bartelmann}.  

\begin{figure}
\includegraphics[height=84mm, angle=-90]{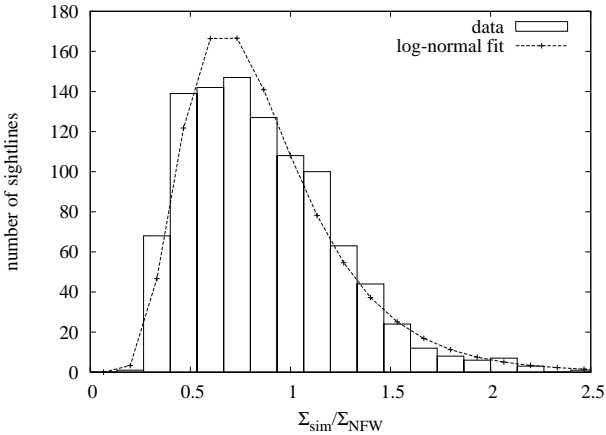}
\caption{
A typical log-normal fit of the scatter. The sightlines have been taken from the halo shown in Fig. \ref{sigma_scatter}, from an impact radius range of $0.7 < r_{\rm imp} / r_{\rm vir} < 0.8$. The surface mass densities are divided in 30 bins, of which only 18 bins containing the majority of the sightlines are visible here. The log-normal distribution is fitted to the frequency of the sightlines in these 30 bins. All fits in our analysis are done accordingly.
}
\label{lnfit}
\end{figure}
\section{A simple correction scheme for smooth and spherical dark matter haloes}

\subsection{Fitting the scatter}

Figure 2 shows the surface mass density scatter of a typical cluster, with its subhaloes removed, and the scatter is seen to follow the NFW profile quite well. For some haloes, there are some strong deviations within the inner 5 \% of the virial radius, and for the majority, a weak, continuous offset trend near the virial radius is seen. The problems of the NFW profile in the core is a known issue \citep{Navarro et al. 2004}. The scatter in surface mass density around the mean is roughly log-gaussian, as will be seen.

The scatter seen in Figure \ref{sigma_scatter} is a consequence of the fact that the haloes are not spherically symmetric -- if they were, all the measured points would land on the NFW profile curve. Subhaloes, when not excluded as in Figure \ref{sigma_scatter}, cause a small number of sightlines to produce even higher surface mass density values than the ones seen in Figure \ref{sigma_scatter}.

For comparing the surface mass densities on sightlines from different haloes (with different virial radii and masses), the impact parameter is transformed to a unitless variable $x_{\rm imp} = r_{\rm imp} / r_{\rm vir}$.

The amount by which each $\Sigma_{\rm SL}(x_{\rm imp})$ deviates from the NFW profile measures how much the shape of the halo deviates from a spherically symmetric analytical model at a specific sightline. This is why the surface mass density is transformed relative to the NFW profile of each halo as follows:
\begin{equation}
S(x_{\rm imp}) = \frac{\Sigma_{\rm SL}(x_{\rm imp})}{\Sigma_{\rm NFW}(x_{\rm imp}c_{\rm vir}, c_{\rm vir})},
\end{equation}
where $c_{\rm vir} = r_{\rm vir}/a_1$ and $\Sigma_{\rm NFW}(x, c)$ is the function in Eq. \ref{tnfw}. As a result of all this, the impact parameter has been scaled to the units of the virial radius and the measured surface mass density to the units of the NFW value.

We divide the measurements in 10 radial bins (one example bin is shown by the vertical lines in Fig. \ref{sigma_scatter}). The width of each bin is $\Delta x_{\rm imp} = 0.1$ and each bin contains 1000 measurements. These radial bins are further divided in 30 bins in surface density, which typically contain a maximum of $\sim 200$ measurements in a single bin (see Fig. \ref{lnfit}).

Once the measurements points are binned, we fit log-normal distributions to the $S(x_{\rm imp})$ distribution, for each radial bin, individually for each halo. The fit function is
\begin{equation}
f(S; \mu, \sigma) = \frac{e^{-({\rm ln}(S)-\mu)^2/(2\sigma^2)}}{S\sigma\sqrt{2\pi}}.
\end{equation}

Repeating the fit procedure for all haloes in the sample allows us to derive the best fit parameters as a function of impact parameter and the halo redshift: $\sigma(x_{\rm imp}, z)$ and $\mu(x_{\rm imp}, z)$. Essentially, $\sigma$ measures the width of the scatter and $\mu$ measures the mean deviation from the NFW fit profile in ln($S$)-space. Because each halo is divided in 10 impact parameter bins, the total number of log-normal fits (and both fit parameters) is $336 \times 10 = 3360$.

\subsection{Finding dependencies to $z$ and $x_{\rm imp}$ in the log-normal parameters}

Finally, we searched for trends in the $\sigma(x_{\rm imp}, z, M_{\rm vir})$ and $\mu(x_{\rm imp}, z, M_{\rm vir})$ data for constructing an analytical description of our measurements. As the analysis progressed, we quickly became aware of the fact that the log-normal parameters do not correlate with the virial mass of the halo as strongly as with the other two variables. Thus, we reduced our analytical description to functions $\sigma(x_{\rm imp}, z)$ and $\mu(x_{\rm imp}, z)$ for the galaxy and cluster sized haloes separately. We also give the analytical description with and without subhaloes, for both mass classes, which then makes the number of our final analytical descriptions four. All four descriptions are expressed by changing parameters within the following function forms:
\begin{equation}
\sigma(x_{\rm imp}, z) = P_0(x_{\rm imp}) + z P_1(x_{\rm imp}),
\label{s_function}
\end{equation}
\begin{equation}
\mu(x_{\rm imp}, z) = Q_0(x_{\rm imp}) + z Q_1(x_{\rm imp}),
\label{m_function}
\end{equation}
where both $P_i(x_{\rm imp})$ and $Q_i(x_{\rm imp})$ are second order polynomials:
\begin{equation}
P_i(x_{\rm imp}) = p_{i2} x_{\rm imp}^2 + p_{i1} x_{\rm imp} + p_{i0},
\end{equation}
\begin{equation}
Q_i(x_{\rm imp}) = q_{i2} x_{\rm imp}^2 + q_{i1} x_{\rm imp} + q_{i0}.
\end{equation}

\begin{figure}
\includegraphics[height=84mm, angle=-90]{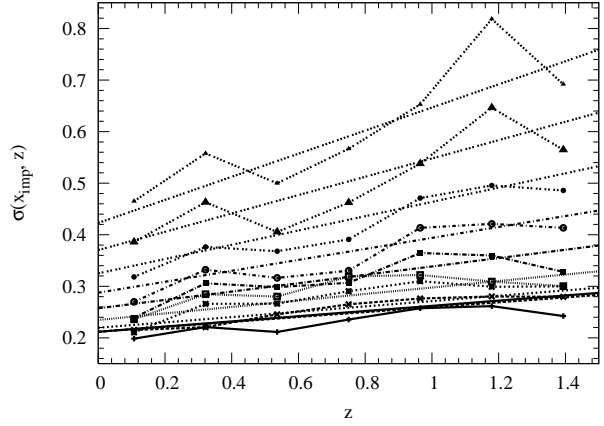}
\caption{
An example of the best fit for the averaged $\sigma(x_{\rm imp}, z)$ fit parameters. Included in this fit are the galaxy sized haloes without subhaloes. Each polyline represents data from a single radial bin and is fitted with the linear function in Equation \ref{s_function}. $\sigma(x_{\rm imp}, z)$ increases with growing impact parameter at a given redshift. The {\it highest} line is the fit for the most distant radial bin $\bar{x}_{\rm imp} = 0.95$, where $0.9 < x_{\rm imp} < 1.0$.
}
\label{s_fit}
\end{figure}

\begin{figure}
\includegraphics[height=84mm, angle=-90]{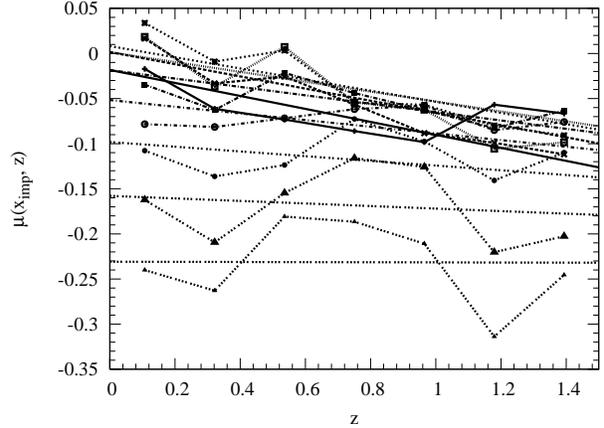}
\caption{
An example of the best fit for the averaged $\mu(x_{\rm imp}, z)$ fit parameters. The halo sample is the same as in Figure \ref{s_fit}. Each polyline represents data from a single radial bin and is fitted with the linear function in Equation \ref{m_function}. The absolute value of $\mu(x_{\rm imp}, z)$ increases with growing impact parameter at a given redshift. The {\it lowest} line is the fit for the most distant radial bin $\bar{x}_{\rm imp} = 0.95$, where $0.9 < x_{\rm imp} < 1.0$.
}
\label{m_fit}
\end{figure}

Examples of the analytical descriptions of $\sigma(x_{\rm imp}, z)$ and $\mu(x_{\rm imp}, z)$ with the actual data can be seen in Figures \ref{s_fit} and \ref{m_fit}. The constants for all polynomials are given in Tables \ref{s_result_table} and \ref{m_result_table}. Before the analytical versions of $\sigma(x_{\rm imp}, z)$ and $\mu(x_{\rm imp}, z)$ are fitted to the log-normal parameters, the data are averaged over bins with a width of $\Delta z \sim 0.2$ to reduce noise. Also, we had to drop the inner 5 \% of the data ($x_{\rm imp} < 0.05$) because within this region, the NFW fit fails to follow the data correctly in a significant number of cases. This is because of our choice of fit weighting (see Section \ref{analytical_haloes}).

Equations \ref{s_function} and \ref{m_function} quantify the distribution of the surface mass density of a generalized galaxy or cluster sized CDM halo at any impact radius and redshift within the ranges we have used. Because of the log-normal distribution, the geometric mean and standard deviations at a given $x_{\rm imp}$ and $z$ values in $S(x_{\rm imp}, z)$-space are
\begin{equation} \label{mu_s}
\mu_S(x_{\rm imp}, z) = e^{\mu(x_{\rm imp}, z)}
\end{equation}
\begin{equation} \label{sigma_s}
\sigma_S(x_{\rm imp}, z) = e^{\sigma(x_{\rm imp}, z)},
\end{equation}
and the upper and lower 1-$\sigma$ limits in $S(x_{\rm imp}, z)$-space are
\begin{equation} \label{sigma_s+}
\sigma_S^+(x_{\rm imp}, z) = e^{(\mu(x_{\rm imp}, z) + \sigma(x_{\rm imp}, z))}
\end{equation}
\begin{equation} \label{sigma_s-}
\sigma_S^-(x_{\rm imp}, z) = e^{(\mu(x_{\rm imp}, z) - \sigma(x_{\rm imp}, z))}.
\end{equation}

We strongly recommend that the analytical function form which {\it includes} subhaloes is used with care. This is because subhaloes can have unwanted effects to the log-normal fitting procedure. The preferred way of using the derived analytical description is to use the version which does not include subhaloes and then add the subhalo scatter afterwards if really necessary (the statistical subhalo contribution to surface mass density is small). A more reliable estimate of the subhalo contribution can be acquired by using the known subhalo mass and distribution functions, which have been studied in detail by e.g. \citet{Gao et al.}. 

\begin{table}
\begin{center}
\caption{
The constants for polynomials $P_0$ and $P_1$, which make up the analytical description for $\sigma(x_{\rm imp}, z)$ (Eq. \ref{s_function}). The samples are coded as follows. G-ns: Galaxy sized haloes, no subhaloes. G-ws: Galaxy sized haloes, with subhaloes. C-ns: Cluster sized haloes, no subhaloes. C-ws: Cluster sized haloes, with subhaloes.
}
\begin{tabular}{lrrr}
\hline
Sample & $p_{00}$ & $p_{01}$ & $p_{02}$ \\
       & $p_{10}$ & $p_{11}$ & $p_{22}$ \\
\hline \hline
G-ns   & 0.217 & $-$0.0692 & 0.305 \\
       & 0.0514 & $-$0.0889 & 0.289 \\
\hline
G-ws   & 0.211 & $-$0.0659 & 0.380 \\
       & 0.0424 & 0.199 & $-$0.00372 \\
\hline
C-ns   & 0.272 & 0.0355 & 0.299 \\
       & 0.0386 & $-$0.121 & 0.269 \\
\hline
C-ws   & 0.287 & 0.0861 & 0.266 \\
       & 0.0260 & $-$0.00533 & 0.247 \\
\hline
\label{s_result_table}
\end{tabular}
\end{center}
\end{table}

\begin{table}
\begin{center}
\caption{
The constants for polynomials $Q_0$ and $Q_1$, which make up the analytical description for $\mu(x_{\rm imp}, z)$ (Eq. \ref{m_function}). The samples are coded as in Table \ref{s_result_table}.
}
\begin{tabular}{lrrr}
\hline
Sample & $q_{00}$ & $q_{01}$ & $q_{02}$ \\
       & $q_{10}$ & $q_{11}$ & $q_{22}$ \\
\hline \hline
G-ns   & $-$0.0322 & 0.286 & $-$0.526 \\
       & $-$0.0739 & 0.0406 & 0.0376 \\
\hline
G-ws   & $-$0.0612 & 0.254 & $-$0.530 \\
       & $-$0.060 & 0.0365 & $-$0.135 \\
\hline
C-ns   & $-$0.0333 & 0.148 & $-$0.433 \\
       & $-$0.0746 & $-$0.00968 & 0.0296 \\
\hline
C-ws   & $-$0.0739 & 0.218 & $-$0.593 \\
       & $-$0.0198 & $-$0.144 & 0.122 \\
\hline
\label{m_result_table}
\end{tabular}
\end{center}
\end{table}

\begin{table}
\begin{center}
\caption{
The minimum and maximum geometric standard deviations and mean offsets from NFW in $S(x_{\rm imp}, z)$-space (see Equations \ref{mu_s} and \ref{sigma_s}). The samples are coded as in Table \ref{s_result_table}. For example, for galaxy sized haloes, with subhaloes, at $z = 1.5$ and with impact parameter $x_{\rm imp} = 1.0$, the 1-$\sigma$ upper limit in surface mass density in units of the NFW model value is $e^{\mu + \sigma} = 0.56 \times 2.41 = 1.35$, and the lower limit is $e^{\mu - \sigma} = 0.56 / 2.41 = 0.23$. Note that the $S(x_{\rm imp}, z)$ distribution is not symmetric but log-normal.
}
\begin{tabular}{lllll}
\hline
Function & G-ns & G-ws & C-ns & C-ws \\ 
\hline \hline
$\sigma_S(0.0, 0.0)$ & 1.24 & 1.23 & 1.32 & 1.33 \\
$\sigma_S(0.0, 1.5)$ & 1.34 & 1.32 & 1.39 & 1.38 \\
$\sigma_S(1.0, 0.0)$ & 1.57 & 1.69 & 1.83 & 1.89 \\
$\sigma_S(1.0, 1.5)$ & 2.29 & 2.41 & 2.42 & 2.82 \\
\hline
$\mu_S(0.0, 0.0)$ & 0.97 & 0.95 & 0.97 & 0.93 \\
$\mu_S(0.0, 1.5)$ & 0.87 & 0.86 & 0.87 & 0.90 \\
$\mu_S(1.0, 0.0)$ & 0.76 & 0.72 & 0.73 & 0.64 \\
$\mu_S(1.0, 1.5)$ & 0.77 & 0.56 & 0.67 & 0.60 \\
\hline
\label{result_table}
\end{tabular}
\end{center}
\end{table}

\subsection{Comparing the model to the data} \label{comparing_model}

We confirm the reliability of our fit procedures, the choice of fit equations and our analytical description of the data by comparing the $\Sigma_{\rm SL}(x_{\rm imp})$-values to the predictions of the model. This is done for all four models by measuring the frequency of the following value:
\begin{equation}
T(x_{\rm imp}, z) = \frac{{\rm ln}(S(x_{\rm imp})) - \sigma(x_{\rm imp}, z)}{\mu(x_{\rm imp}, z)}.
\end{equation}

\begin{figure}
\includegraphics[height=84mm, angle=-90]{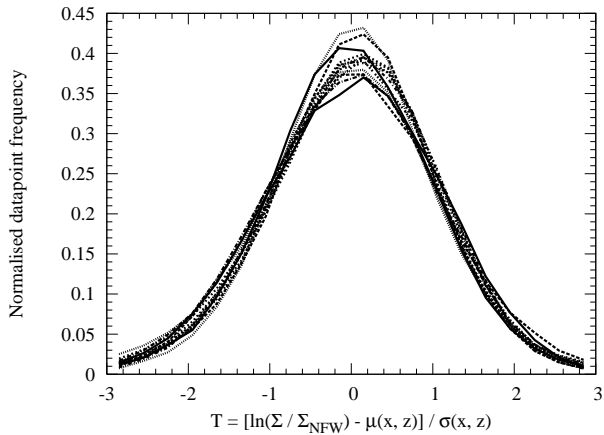}
\caption{
Here one of our analytical models (galaxy sized haloes, no subhaloes) is compared against the data. Different curves represent different redshifts, and there is no trend between them. All curves are close to the standard normal distribution (which peaks at $\sim 0.4$), which means that our model is able to describe the distribution of the measured values quite well. See Section \ref{comparing_model} for details.}
\label{model_vs_data}
\end{figure}

The probability density of $T(x_{\rm imp}, z)$ should follow a standard normal distribution for all redshifts if the model is reliable. As shown by the example in Figure \ref{model_vs_data}, this is the case when all the 10,000 $\times 130$ sightlines for the galaxy sized haloes (without subhaloes) are considered at all redshifts.

The fact that the resulting distributions are centered on zero tells us that $\mu(x_{\rm imp}, z)$ is reconstructed correctly. The standard deviations of the distributions are close to unity, which again tells us that $\sigma(x_{\rm imp}, z)$ is a fair estimate of the behaviour of the data at all impact radii and redshifts. The test was equally successful for galaxy and cluster sized haloes, with and without subhaloes as the test shown in Figure \ref{model_vs_data}.

\section{Impact on the MACHO optical depth calculations}

MACHOs (Massive Astrophysical Compact Halo Objects) represent one class of dark matter candidates, which may be detected through gravitational microlensing effects as they pass through the line of sight to distant light sources. Although the MACHO acronym was originally invented with baryonic objects like faint stars and stellar remnants in mind, several non-baryonic dark matter candidates like axion aggregates \citep{Membrado}, mirror matter objects \citep{Mohapatra & Teplitz}, primordial black holes \citep{Green}, quark nuggets \citep{Chandra & Goyal}, preon stars \citep{Hansson & Sandin} and scalar dark matter miniclusters \citep{Zurek et al.} can also manifest themselves in this way. While non-baryonic MACHOs could in principle account for a substantial fraction of the dark matter, microlensing searches based on observations of light sources in the local Universe suggest a MACHO dark matter fraction of $\sim 20$ \% \citep[e.g.][]{Alcock et al.,Calchi Novati et al.}. Low-redshift microlensing observations are only able to detect the densest, most compact MACHOs at a given mass. This does not necessarily represent a robust upper limit on the relative importance of MACHOs, and high-redshift observations will be required to settle the issue (see \citealt{Zackrisson & Riehm} for a more detailed review).

While difficult to directly relate to observational quantities, the microlensing optical depth $\tau$ is often used for estimates of the viability and relevance of different microlensing scenarios. Formally, the microlensing optical depth represents the average number of lenses along a random line of sight. Under the assumption that the lenses do not overlap along the sightline, it also represents the fraction of sky that is covered by regions in which a point source will be microlensed. In the limit of small $\tau$ the optical depth can therefore directly be used as an estimate of the microlensing probability. At higher $\tau$, this interpretation does however break down because of overlapping microlenses. Here, the optical depth will be used to discuss the impact of surface mass density variations due to clumpy and non-spherical CDM haloes on MACHO microlensing calculations. In the following, we assume the spatial distribution of MACHOs to follow that of the CDM, as would be expected if they are non-baryonic. We caution that these results do not necessarily apply to baryonic MACHOs, since such objects may have a spatial distribution that is substantially different from that of the overall dark matter \citep[as illustrated by population III stars; e.g.][]{Scannapieco et al.}.

In the extreme case of having all CDM in the form of MACHOs, the MACHO optical depth along a sightline passing through a single halo can be approximated by: 
\begin{equation}
\tau=\frac{\Sigma_{CDM}}{\Sigma_\mathrm{c}},
\label{optdepth_halo}
\end{equation}
where $\Sigma_\mathrm{CDM}$ is the CDM surface mass density of this particular sightline and $\Sigma_\mathrm{c}$ is the critical surface mass density for lensing. The latter is given by:
\begin{equation}
\Sigma_\mathrm{c}=\frac{c^2}{4\pi G} \frac{D_\mathrm{os}}{D_\mathrm{ls} D_\mathrm{ol}},
\end{equation}
where $D_\mathrm{ol}$, $D_\mathrm{ls}$ and $D_\mathrm{os}$ are the angular-size distances from observer to lens, lens to source and observer to source, respectively. In the case when only a fraction of the CDM is the form of MACHOs, and the rest is in the form of a smooth component, the optical depth can instead be estimated using \citep[e.g.][where an analogous case with stellar microlenses in a smooth matter distribution is considered]{Wyithe & Turner}:
\begin{equation}
\tau=\frac{f_\mathrm{MACHO}\Sigma_\mathrm{CDM}}{|\Sigma_\mathrm{c}-(1-f_\mathrm{MACHO})\Sigma_\mathrm{CDM}|},
\end{equation}
where $f_\mathrm{MACHO}$ represents the MACHO fraction. 

In situations where the sightline is dominated by a single halo of (approximately) known mass, and the impact parameter of this sightline can be estimated (in units of  $r_\mathrm{vir}$), the fitting formulas presented in Sect. 5 may be directly applied to assess the mean and variance in the expected MACHO optical depth derived from a spherical model for the foreground halo. This situation occurs when estimating the contribution from non-baryonic MACHOs in the dominating halo to the total microlensing optical depth towards a gravitational arcs in a galaxy cluster, or an individual image of a strongly lensed quasar. As described in Section 5, the error on the optical depth due to triaxiality can in this situation easily amount to factor of $\sim 2$ (assuming Eq. \ref{optdepth_halo} for the optical depth). Note that this will be compounded by the surface mass density error coming from the uncertainty in the mass and concentration parameter of the spherical model for the foreground halo. In situations where many light sources are monitored, and these are projected across a large area of the foreground halo (as in the cluster-quasar and cluster-cluster microlensing monitoring programmes of \citealt{Totani} and \citealt{Tadros et al.}, respectively), care must be exercised when using the formulae presented in Sect. 5. When averaging over different, but not independent sightlines, the effects of triaxiality may be substantially diminished, and the optical depth error derived from these formulae should be considered a conservative upper limit.

In many cases, however, the impact parameter of the dominating halo is not known, and there may be more than one halo giving significant contributions to the optical depth of a given sightline. This happens when light sources, especially at high redshift, are randomly selected without reference to any foreground object. In this case, the average microlensing optical depth is often computed using: 
\begin{equation}
\bar{\tau} = \frac{3H_0\Omega_\mathrm{MACHO}}{2D_\mathrm{os}}\int^{z_\mathrm{s}}_{0} \frac{(1+z)^2 D_\mathrm{ls} D_\mathrm{ol} \mathrm{d}z}{\sqrt{\Omega_\mathrm{M}(1+z)^3+\Omega_\Lambda}}, 
\label{optdepth_aver}
\end{equation}
where $z_\mathrm{s}$ is the redshift of the light source studied and $\Omega_\mathrm{MACHO}$ is the cosmological density of MACHOs relative to critical at zero redshift. This estimate, often referred to as the \citet{Press & Gunn} approximation, also assumes a constant comoving number density of MACHOs, i.e. that the MACHO population does not evolve as a function of redshift. Since the matter of the Universe is clustered, one does however expect a certain scatter around this average, since some sightlines will contain more matter (and hence MACHOs) than others. \citet{Zackrisson & Riehm} find, using a model that takes into account the clustering of MACHOs into spherical CDM haloes and subhaloes, that the distibution of MACHO optical depths around $\bar{\tau}$ is reasonably well described by a log-normal function with standard deviation $\sigma_{\ln \tau}(z_\mathrm{s})$. As the number of intervening haloes increases when more and more distant light sources are considered, the sightline-to-sightline scatter, and hence $\sigma_{\ln \tau}(z_\mathrm{s})$, decreases with increasing $z_\mathrm{s}$. In this model, the optical depth scatter is dominated by the different number of haloes along each sightline, combounded by the different masses, concentration parameters and impact parameters for each of these objects. 

Since non-sphericity introduces additional optical depth scatter on top of that produced by the spherical halos, one may expect halo triaxiality to give an significant contribution to $\sigma_{\ln \tau}(z_\mathrm{s})$, but when implementing the $\Sigma_\mathrm{CDM}(r)$ scatter formulae derived here in the \citet{Zackrisson & Riehm} microlensing code, we find the impact of non-sphericitiy to be negligible. The reason for this is that the amplitude of the scatter stemming from triaxiality is relatively small compared to that coming from other aspects of CDM clustering. As can be seen in Figure \ref{sigma_scatter}, the mean surface mass density varies by more than a factor of $\sim 1000$ between the innermost regions of the halo and $r_\mathrm{vir}$, whereas triaxiality generates variations of less than a factor of $\sim 10$ at each impact parameter. This means that, once random halo impact parameters are considered, the resulting MACHO optical depth distribution will be dominated by the scatter introduced by the form of the surface mass density profile, while halo triaxiality will be responsible for only a very slight modification of this distribution. Considering the fact that most sightlines towards high-redshift sources pass inside $r_\mathrm{vir}$ of more than one halo \citep[see][for an estimate of how many]{Zackrisson & Riehm}, and that these are likely to have different masses and concentration parameters, the impact of triaxiality on the optical depth distribution quickly becomes negligibly small. Hence, it can safely be ignored in this situation. This greatly reduces the computational complexity of MACHO microlensing models for high-redshift sources. 

\section{Discussion}

The results presented here do suffer from a number of shortcomings which should be pointed out. The simulations used are dissipationless. In reality, dark matter haloes contain baryons, and the dissipation and feedback associated with these will inevitably affect the overall potential of the system, and thereby the spatial distribution of the CDM. According to current models, baryonic cooling will increase the central density of the CDM \citep[e.g.][]{Gnedin et al.} and also make the halo more spherical \citep{Kazantzidis et al.}. The significance of these effects are, however, still difficult to predict reliably, as the gas dynamical simulations involved still suffer from so-called ``overmerging'' problems \citep[e.g.][]{Balogh et al., Springel & Hernquist}.  

When calculating the surface mass density profiles, we have moreover considered only the matter present within $r_\mathrm{vir}$ of each halo, whereas simulations have shown that galaxy sized CDM haloes extend at least out to 2--3$r_\mathrm{vir}$ \citep{Prada et al.}. We restricted our analysis to $r_\mathrm{vir}$ because it becomes increasingly demanding to separate halo particles from the background the further one wants to extend the analysis. Even in the presented case, we need to separate particles out to $\sim 1.5 r_\mathrm{vir}$ because the smoothed particles extend their influence inside the virial radius region even though they are positioned outside it. We tested our method out to 3$r_\mathrm{vir}$, but the number counts of the halo particles at those distances are too low to produce reliable results. Our halo sample does not have the resolution needed for extending the analysis further than $r_\mathrm{vir}$ safely.

The most significant limitation of this paper is the small number of haloes in our analysis. This is of course due to the limited resolution of the cosmological simulations we had access to. We would like to repeat our analysis with a more complete statistical sample of haloes, which would hopefully confirm our analytical description with smaller error bars.

We also note that our analytical description of the surface mass density is more reliable in the case in which subhaloes are {\it excluded}. This is because subhaloes can introduce significant mass peaks to some radial bins. These peaks can lead to unwanted effects in the log-normal fitting procedure which is designed to handle relatively smooth and continuous mass distributions within a bin. Large subhalos can also disturb the NFW fits, at last in the low density regions. Thus, the use of the models which include subhaloes is discouraged.

\section{Summary}

We have compared a sample of CDM N-body haloes to the smooth, spherically symmetric NFW density profile model in three dimensions. The differences in surface mass density of the haloes and the model are studied, and an analytical description of the differences is constructed. This description can be used to estimate or reproduce the differences between CDM N-body haloes and, in principle, any analytical halo model. It can be used in applications in which the line-of-sight surface mass densities of CDM haloes play an important role, such as microlensing.

Our halo sample consists of 27 {\it independent} CDM haloes at $\sim 10$ redshift snapshots between $0.0 < z < 1.5$. The haloes are extracted from six cosmological simulations with comoving box sizes of 10 $h^{-1}$Mpc, 40 $h^{-1}$Mpc and 64 $h^{-1}$Mpc. The haloes are treated both with and without their subhaloes, and the halo sample is divided in two mass classes, separated by a mass gap at $M \sim 10^{13}$ M$_\odot h^{-1}$. The analytical description is given for all four cases.

We find that the surface mass density of the haloes can deviate from the spherical model considerably. At minimum, with zero impact parameter and redshift, the 1-$\sigma$ limits around the NFW surface mass density are close to $\sigma = \pm 20 \%$ or $\sigma = \pm 30 \%$, depending which haloes are under investigation. At maximum, with impact parameter close to $r_{\rm vir}$ and z = 1.5, the values can be as high as $\sigma_+ = +70 \%$ and $\sigma_- = -80 \%$. The geometric mean of the surface mass density is offset from the NFW predicted value by $-3$ \% to $-44$ \%, depending on the case.

We also find that the departure from the NFW profile is log-normally distributed around the model value. In most cases, the median of the surface mass density of the haloes is slightly lower than predicted by the NFW profile. The variation of the surface mass density around the NFW value grows with increasing impact parameter and redshift. 

As an application, we introduce our analytical description to the optical depth calculations of MACHOs. In this case, we find that the variance in surface mass density due to halo shapes can be overwhelmed by the variance caused by random impact parameters between halos on the same sightline.

\section*{Acknowledgements}
JH acknowledges PhD studentship grants from the Vilho, Yrj\"o and Kalle V\"ais\"al\"a and the Magnus Ehrnrooth foundations, and travel grants from the Otto A. Malmi foundation and the Turku University Foundation (Valto Takala Fund). EZ acknowledges research grants from the Swedish Research Council, the Royal Swedish Academy of Sciences and the Academy of Finland. AK acknowledges funding through the Emmy Noether Programme by the DFG (KN 755/1). CF acknowledges the Academy of Finland. The cosmological simulations were run at the Swinburne University of Technology and at the CSC - Scientific Computing center in Finland.

\label{lastpage}

\begin{thebibliography}{}
\bibitem[\protect\citeauthoryear{Alcock et al.}{2000}]{Alcock et al.}
Alcock C. et al., 2000, ApJ, 542, 281
\bibitem[\protect\citeauthoryear{Balogh et al.}{2001}]{Balogh et al.}
Balogh M. L., Pearce F. R., Bower R. G., Kay S. T., 2001, MNRAS, 326, 1228
\bibitem[\protect\citeauthoryear{Bartelmann \& Weiss}{1994}]{Bartelmann & Weiss}
Bartelmann M., Weiss A., 2004, A\&A, 287, 1
\bibitem[\protect\citeauthoryear{Bartelmann}{1996}]{Bartelmann} 
Bartelmann M., 1996, A\&A, 313, 697 
\bibitem[\protect\citeauthoryear{Bertschinger \& Gelb}{1991}]{Bertschinger & Gelb}
Bertschinger E., Gelb J. M., 1991, CP 5, 164
\bibitem[\protect\citeauthoryear{Calchi Novati et al.}{2005}]{Calchi Novati et al.}
Calchi Novati S. et al. 2005, A\&A, 443, 911
\bibitem[\protect\citeauthoryear{Chae}{2003}]{Chae}
Chae K.-H., 2003, MNRAS, 346, 746
\bibitem[\protect\citeauthoryear{Chandra \& Goyal}{2000}]{Chandra & Goyal}
Chandra D., Goyal A., 2000, Phys. Rev. D, 62, 63505
\bibitem[\protect\citeauthoryear{Dalal et al.}{Dalal, Holder \& Hennawi}{2004}]{Dalal et al.}
Dalal N., Holder G., Hennawi J. F., 2004, ApJ, 609, 50
\bibitem[\protect\citeauthoryear{Davis et al.}{1985}]{Davis et al.}
Davis M., Efstathiou G., Frenk C. S., White S. D. M., 1985, ApJ 292, 371
\bibitem[\protect\citeauthoryear{Evans \& Hunter}{2002}]{Evans & Hunter}
Evans N. W., Hunter C., 2002, ApJ, 575, 68
\bibitem[\protect\citeauthoryear{Frenk et al.}{1988}]{Frenk et al.}
Frenk C. S., White S. D. M., Davis M., Efstathiou G., 1988, ApJ 327, 507
\bibitem[\protect\citeauthoryear{Gao et al.}{2004}]{Gao et al.} 
Gao L., White S.~D.~M., Jenkins A., Stoehr F., Springel V., 2004, MNRAS, 355, 819
\bibitem[\protect\citeauthoryear{Gnedin et al.}{2004}]{Gnedin et al.}
Gnedin O. Y., Kravtsov A. V., Klypin A. A., Nagai D., 2004, ApJ 616, 16
\bibitem[\protect\citeauthoryear{Gill et al.}{2004}]{Gill et al.}
Gill S. P. D., Knebe A., Gibson B. K., 2004, MNRAS, 351, 399
\bibitem[\protect\citeauthoryear{Golse \& Kneib}{2002}]{Golse & Kneib}
Golse G., Kneib J.-P., 2002, A\&A, 390, 821
\bibitem[\protect\citeauthoryear{Green}{2000}]{Green}
Green A. M., 2000, ApJ, 537, 708
\bibitem[\protect\citeauthoryear{Gunnarsson}{2004}]{Gunnarsson}
Gunnarsson C., 2004, JCAP 03, 2 
\bibitem[\protect\citeauthoryear{Hansson \& Sandin}{2005}]{Hansson & Sandin} 
Hansson J., Sandin F., 2005, Phys. Lett. B, 616, 1
\bibitem[\protect\citeauthoryear{Hennawi et al.}{2007}]{Hennawi et al.}
Hennawi J. F., Dalal N., Bode P., Ostriker J. P., 2007, ApJ 654, 714
\bibitem[\protect\citeauthoryear{Hockney \& Eastwood}{1981}]{hockney}
Hockney R. W., Eastwood J. W., 1981, Computer Simulation Using Particles, McGraw-Hill, New York, p. 144
\bibitem[\protect\citeauthoryear{Holopainen et al.}{2006}]{Holopainen et al.} 
Holopainen J., Flynn C., Knebe A., Gill S. P., Gibson B. K., 2006, MNRAS 368, 1209
\bibitem[\protect\citeauthoryear{Jing \& Suto}{2002}]{Jing & Suto}
Jing Y. P., Suto Y., 2002, ApJ, 574, 538 
\bibitem[\protect\citeauthoryear{Kazantzidis et al.}{2004}]{Kazantzidis et al.}
Kazantzidis S., Kravtsov A. V., Zentner A. R., Allgood B., Nagai D., Moore B., 2004, ApJ 611, L73
\bibitem[\protect\citeauthoryear{Klypin \& Holtzman}{1997}]{Klypin & Holtzman}
Klypin A. A., Holtzman J., 1997, astro-ph 9712217
\bibitem[\protect\citeauthoryear{Knebe, Green~\& Binney}{2001}]{MLAPM}
Knebe A., Green A., Binney J., 2001, MNRAS 325, 845
\bibitem[\protect\citeauthoryear{Knebe \& Wiessner}{2006}]{KnebeWiessner}
Knebe A., Wiessner V., 2006, PASA 23, 125 
\bibitem[\protect\citeauthoryear{Kochanek}{1996}]{Kochanek}
Kochanek C. S., 1996, ApJ, 473, 595
\bibitem[\protect\citeauthoryear{Moore et al.}{1999}]{Moore et al.}
Moore B., Ghigna S., Governato F., Lake G., Quinn T., Stadel J., Tozzi P., 1999, ApJ, 524, l19
\bibitem[\protect\citeauthoryear{Membrado}{1998}]{Membrado}
Membrado M., 1998, MNRAS, 296, 21
\bibitem[\protect\citeauthoryear{Mohapatra \& Teplitz}{1999}]{Mohapatra & Teplitz}
Mohapatra R. N., Teplitz V. L., 1999, Phys. Lett. B., 462, 302
\bibitem[\protect\citeauthoryear{NFW}{Navarro, Frenk \& White}{1996}]{NFW}
Navarro J. F., Frenk C. S., White S. D. M., 1996, ApJ, 462, 563
\bibitem[\protect\citeauthoryear{Navarro et al.}{2004}]{Navarro et al. 2004}
Navarro J. F. et al., 2004, MNRAS, 349, 1039 
\bibitem[\protect\citeauthoryear{Oguri \& Keeton}{2004}]{Oguri & Keeton}
Oguri M., Keeton C. R., 2004, ApJ, 610, 663 
\bibitem[\protect\citeauthoryear{Oguri}{2005}]{Oguri b}
Oguri M., 2005, MNRAS 367, 1241
\bibitem[\protect\citeauthoryear{Prada et al.}{2006}]{Prada et al.}
Prada F., Klypin A. A., Simonneau E., Betancort-Rijo J., Patiri S., Gottl\"ober S., Sanchez-Conde M. A. 2006, ApJ, 645, 1001
\bibitem[\protect\citeauthoryear{Press \& Gunn}{1973}]{Press & Gunn}
Press W. H., Gunn J. E., 1973, ApJ, 185, 397
\bibitem[\protect\citeauthoryear{Profumo \& Sigurdson}{2007}]{Profumo & Sigurdson}
Profumo S., Sigurdson K. 2007, Phys.Rev. D75, 023521
\bibitem[\protect\citeauthoryear{Primack}{2003}]{Primack}
Primack J. R., 2003, Nuclear Physics B Proceedings Supplements, 124, 3
\bibitem[\protect\citeauthoryear{Scannapieco et al.}{2006}]{Scannapieco et al.}
Scannapieco E., Kawata D., Brook C. B., Schneider R., Ferrara A., \& Gibson B. K., 2006, ApJ, 653, 285 
\bibitem[\protect\citeauthoryear{Seljak \& Holz}{1999}]{Seljak & Holz}
Seljak U., Holz D. E., 1999, A\&A 351, L10
\bibitem[\protect\citeauthoryear{Springel \& Hernquist}{2002}]{Springel & Hernquist}
Springel V., Hernquist L., 2002, MNRAS 333, 649
\bibitem[\protect\citeauthoryear{Suto, Cen \& Ostriker}{1992}]{SCO}
Suto Y., Cen R., Ostriker J., 1992, ApJ 395, 1
\bibitem[\protect\citeauthoryear{Tadros et al.}{Tadros, Warren \& Hewett}{2001}]{Tadros et al.}
Tadros H., Warren S.,  Hewett P., 2001, New Astron. Rev., 42, 115
\bibitem[\protect\citeauthoryear{Tormen}{1997}]{Tormen97}
Tormen G., 1997, MNRAS 290, 411
\bibitem[\protect\citeauthoryear{Totani}{2003}]{Totani}
Totani T. 2003, ApJ, 586, 735
\bibitem[\protect\citeauthoryear{Warnick \& Knebe}{2006}]{WK}
Warnick K., Knebe A., 2006, MNRAS 369, 1253
\bibitem[\protect\citeauthoryear{Warnick, Knebe \& Power}{2007a}]{WKPI}
Warnick K., Knebe A., Power C., 2007a, MNRAS submitted
\bibitem[\protect\citeauthoryear{Warnick, Knebe \& Power}{2007b}]{WKPII}
Warnick K., Knebe A., Power C., 2007b, MNRAS submitted
\bibitem[\protect\citeauthoryear{Weinberg, Hernquist \& Katz}{1997}]{WHK}
Weinberg D., Hernquist L., Katz N., 1997, ApJ 477, 8
\bibitem[\protect\citeauthoryear{Wyithe \& Turner}{2002}]{Wyithe & Turner}
Wyithe J.S.B., Turner E.L., 2002, ApJ 567, 18
\bibitem[\protect\citeauthoryear{Zackrisson \& Riehm}{2007}]{Zackrisson & Riehm}
Zackrisson E., \& Riehm T., 2007, A\&A, accepted (arXiv:0709.1571)
\bibitem[\protect\citeauthoryear{Zhao}{1996}]{Zhao}
Zhao H. S., 1996, MNRAS, 278, 488
\bibitem[\protect\citeauthoryear{Zurek et al.}{Zurek, Hogan \& Quinn}{2006}]{Zurek et al.}
Zurek K. M., Hogan C. J., Quinn T. R., 2006, astro-ph/0607341
\end{thebibliography}
\end{document}